\documentclass{aa}  
\usepackage{graphicx}
\usepackage{txfonts}
\usepackage{natbib}
\bibpunct{(}{)}{;}{a}{}{,}
\begin{document}
   \title{Parametrization of C-shocks. 
          Evolution of the Sputtering of Grains}


   \author{I. Jim\'enez-Serra\inst{1,2}, P. Caselli\inst{2,3},
          J. Mart\'{\i}n-Pintado\inst{1} 
          \and T. W. Hartquist\inst{2}}

   \offprints{izaskun@damir.iem.csic.es}

   \institute{Departamento de Astrof\'{\i}sica Molecular e Infrarroja,
             Instituto de Estructura de la Materia (CSIC),
             C/ Serrano 121, E-28006 Madrid, Spain
         \and
              School of Physics and Astronomy, University of Leeds LS2
	      9JT, Leeds, United Kingdom
         \and 
             INAF-Osservatorio Astrofisico di Arcetri, Largo E. Fermi 5,
	     I-50125 Firenze, Italy}

   \date{Received September 15, 1996; accepted March 16, 1997}

 
  \abstract
   {The detection of a narrow SiO line emission toward the young shocks of
  the L1448-mm outflow has been interpreted as a signature of the
  magnetic precursor of C-shocks. In contrast with the very low SiO 
  abundances ($\leq$10$^{-12}$) derived from the ambient gas, 
  the narrow SiO emission in the precursor component 
  at almost ambient velocities reveals enhanced
  SiO abundances of $\sim$10$^{-11}$. This enhancement has been
  proposed to be produced by the sputtering of the grain mantles at
  the very first stages of C-shocks. However,
  modelling of the sputtering of grains 
  has usually averaged the SiO abundances
  over the dissipation region of C-shocks, which cannot
  explain the recent observations.} 
   {To model the evolution of the gas phase abundances of molecules
  like SiO, CH$_3$OH and H$_2$O, produced by the sputtering of 
  the grain mantles and cores as the shock propagates through the
  ambient gas. We consider different initial gas densities and 
  shock velocities.}
   {We propose a parametric model to describe the physical structure
  of C-shocks as a function of time. Using the known sputtering yields
  for water mantles (with other minor constituents like silicon
  and CH$_3$OH) and olivine cores by collisions with H$_2$, He, 
  C, O, Si, Fe and CO, we follow the evolution of the
  abundances of silicon, CH$_3$OH and H$_2$O ejected from grains along
  the evolution of the shock.}   
  {The evolution of the abundances of the sputtered silicon, CH$_3$OH
  and H$_2$O shows that CO seems to be the most 
  efficient sputtering agent in low velocity shocks. The velocity
  threshold for the sputtering of silicon from the grain mantles is appreciably
  reduced (by 5-10$\,$km$\,$s$^{-1}$) by CO compared to other models. 
  The sputtering by CO can generate SiO abundances of $\sim$10$^{-11}$
  at the early stages of low velocity shocks, consistent with those
  observed in the magnetic precursor component of L1448-mm. 
  Our model satisfactorily reproduce the progressive enhancement of 
  SiO, CH$_3$OH and H$_2$O observed in this outflow, suggesting 
  that this enhancement may be due to the propagation of two 
  shocks with $v_s$=30$\,$km$\,$s$^{-1}$ and $v_s$=60$\,$km$\,$s$^{-1}$ 
  coexisting within the same region.} 
  {Our simple model can be used to estimate the time
  dependent evolution of the abundances of molecular shock tracers
  like SiO, CH$_3$OH, H$_2$O or NH$_3$ in very young molecular outflows.}


   \authorrunning{Jim\'enez-Serra et al.}
   \titlerunning{Evolution of the Sputtering of Grains}
   \maketitle
%

\section{Introduction}

In young molecular outflows, it is expected that changes in the
molecular emission could be observed 
due to the propagation of shocks into the 
ambient material. So far, the L1448-mm outflow is the only object 
where time variability of the SiO emission in the high velocity jet 
has been detected, indicating the presence of very young 
shocks \citep[][]{gir01}. 

It is well known that silicon is heavily depleted onto the grain
mantles and grain cores in 
the quiescent gas of molecular dark clouds like TMC-1, L183 and L1448 
(SiO abundance of $\leq$10$^{-12}$; Ziurys, Friberg \& Irvine 1989; 
Mart\'{\i}n-Pintado, Bachiller \& Fuente 1992; Requena-Torres et
al. 2007). The detection of large SiO abundances in regions 
with outflow activity is therefore a clear indicator of the destruction of 
dust grains by the interaction of magnetohydrodynamic
(MHD) shock waves (or C-shocks) with the ambient gas
\citep[][]{mar92,flo96,cas97}. 

The typical SiO abundances measured in the high velocity gas of 
young molecular outflows like in L1448-mm 
\citep[][]{mar92} are of
$\geq$10$^{-6}$, which implies an enhancement by more than 6 orders of
magnitude with respect to the SiO abundances measured in the quiescent
gas. The recent detection of very narrow SiO emission at almost ambient
velocities toward this outflow has been proposed to be produced by the 
magnetic precursor of C-shocks \citep[][]{jim04}. The SiO abundance
of $\sim$10$^{-11}$ for this narrow emission 
clearly contrasts with the large SiO enhancement found in 
the high velocity postshock gas, and with the much lower SiO abundance of
the quiescent material. 

Toward the young shocks of the L1448-mm outflow, 
\citet{jim05} have also reported an 
evolutionary trend of the SiO and CH$_3$OH abundances \citep[methanol
  is the most abundant molecule after H$_2$O in the grain mantles;][]{tie87} 
to be enhanced from the ambient gas to the moderate
velocity component, as if the grain mantles would
have been progressively eroded by the recent interaction
of low velocity shocks. 

Modelling of C-shocks that includes only the sputtering of grain
cores, shows that an appreciable fraction of silicon material starts
to be ejected from grains for 
$v_s\geq$25$-$30$\,$km$\,$s$^{-1}$ \citep{flo96,cas97,may00}. Although these 
models predict SiO abundances consistent with those observed 
in the postshock gas ($\sim$10$^{-8}$-10$^{-7}$), 
the sputtering of SiO from the cores cannot reproduce the lower 
SiO abundances of $\sim$10$^{-11}$ found in the narrow precursor
component of L1448-mm. 

Calculations of the sputtering yield of silicon by heavy atoms like 
C, O, Si and Fe on SiO$_2$ and olivine (MgFeSiO$_4$) cores show that, 
despite the low relative abundances of these species with respect to 
H$_2$ and He in dark clouds, these heavy particles can dominate the 
sputtering of grains at low shock velocities \citep{fie97,may00}. 
Furthermore, abundant molecules like CO could also
play an important role in the sputtering of dust grains since 
these species can sputter like atoms of equivalent mass for low impact
velocities \citep{may00}. Considering that silicon could be a minor
constituent of the mantles, their sputtering by these heavy species in low
velocity shocks could efficiently erode them generating
the SiO abundances observed for the narrow SiO line emission 
in L1448-mm. Up to date, 
the evolution of the sputtering of grains has not been studied in
detail. The questions of which species 
are the most efficient sputtering agents, 
and which time-scales 
are needed to eject most of the silicon material from grains, 
still remain uncertain.  


In this paper, we present a parametric model of C-shocks 
to describe in detail the time dependent evolution of the molecular
abundances sputtered from grains in low and high velocity shocks. 
This approximation constitutes a powerful tool for 
interpreting the molecular abundances measured in young molecular
outflows. In addition to H$_2$ and He, heavy atoms and molecules have
been also considered as sputtering agents.   
In Sec.$\,$2, we present the approximations used to describe the
steady state profile of the physical structure of C-shocks. 
In Sec.$\,$3, we show the procedure used to determine the sputtering
of the grain mantles and the grain cores. 
In Sec.$\,$4, we present the results of the sputtering of silicon from
grains for several initial gas densities and shock velocities. In
Sec.$\,$5 and 6, we compare the sputtered SiO, CH$_3$OH and H$_2$O 
abundances with those measured in the L1448-mm outflow. The
conclusions are finally summarized
in Sec.$\,$7. 

\section{The C-shock Structure in the Preshock Frame}

We consider a plane-parallel C-shock that propagates through the
quiescent gas with velocity $v_s$. As a first approximation, we 
have assumed steady state profiles for the evolution of the physical
parameters in the shock. The validity of this approximation, 
versus more recent time-dependent modelling of the physical 
structure of C-shocks, will be discussed in detail in 
Sec.$\,$4.1. 

%
\begin{table}
\caption{Initial gas phase abundances of He, C, O, Si, Fe and CO.}              
\label{tab0}
\centering 
\renewcommand{\footnoterule}{}  
\begin{tabular}{cc}     
\hline\hline       
 Element/ & Abundance$^{\mathrm{a}}$ \\ 
 Molecule & [$n$(X)/$n$(H)]           \\
\hline                    
   He & 0.1 \\
   C  & 7.1$\times$10$^{-9}$ \\
   O  & 1.8$\times$10$^{-4}$ \\
   Si & 8.0$\times$10$^{-9}$ \\
   Fe & 3.0$\times$10$^{-9}$ \\
   CO & 1.5$\times$10$^{-4}$ \\
\hline                  
\end{tabular}
\begin{list}{}{}
\item[$^{\mathrm{a}}$] From the low metal case 
for dense cloud chemistry \citep{pra82,her89,gra82}.
\end{list}
\end{table}
%

The initial H$_2$ density and temperature of the ambient cloud are $n_0$ 
and $T_0$, respectively. Since one of the aims of this work is to directly 
compare our results with observations toward the young L1448-mm outflow, it is 
convenient to consider that the velocities of the ion and neutral fluids, 
$v_i$ and $v_n$, are in the frame co-moving with the preshock gas. These 
velocities are approximated by:

   \begin{eqnarray}
      v_{n,i} & = & \left(v_s -v_0\right) - \frac{\left(v_s-v_0\right)}{cosh\left[(z-z_0)/z_{n,i}\right]} \label{vel}
   \end{eqnarray}

where $z$ is the spatial coordinate and the $z_{n}/z_{i}$ 
ratio governs the strength of the velocity decoupling between
the ion and neutral fluids. $z_0$ corresponds to the distance at
which these fluids start to decouple (see Sec.$\,$4.1 for details on how to 
estimate these parameters). An additional velocity, $v_0$, 
also needs to be considered in the equations for $v_n$ and $v_i$ 
in order to avoid infinite compression of the far downstream gas 
(see Eq.$\,$\ref{dens} below). $v_0$ depends on the shock parameters and is defined 
as the final downstream velocity of the ion and neutral fluids 
{\it in the frame of the shock} (in the preshock frame, this final velocity 
would be $v_s-v_0$; see Sec.$\,$4.1). As shown in Appendix A, this velocity 
is tightly linked to the shock and Alfv\'en 
velocities, $v_s$ and $v_A$, through shock jump conditions.

The ion-neutral drift speed $v_d$ is $v_d=$$|v_n-v_i|$, and  
the neutral fluid flow time is calculated as (see 
Eqs.$\,$\ref{trapz} and \ref{f}):  

   \begin{eqnarray}
      t = \int \frac{dz}{v_s-v_n} \label{time}
   \end{eqnarray}

From the principle of mass conservation, the neutral density, $n_n$, is
given by:
   \begin{eqnarray}
      n_n & = & \frac{n_0 v_s}{v_s-v_n} \label{dens}
   \end{eqnarray}
The temperature of the neutral fluid, $T_n$, is approximated 
by a Planck-like function as:
   \begin{eqnarray}
      T_n & = & T_0 + \frac{[a_{T}\,(z-z_0)]^{b_T}}{exp[(z-z_0)/z_{T}]-1} \label{temp}
   \end{eqnarray}
where $b_T$ is an integer, and $a_T$ and $z_T$ are
related to the maximum value of $T_{n}$ ($T_{n,max}$) and the
distance $z_{n,max}$ at which $T_n$ reaches its maximum value.  
The temperature of the ion fluid is calculated by using 
$T_i$$\,$=$\,$$T_n$+$\left(\frac{m\,v_d^2}{3\,k}\right)$. 

%
\begin{table}
\caption{Fractional abundances of H$_2$O, CH$_3$OH and Si/SiO
    assumed for the icy mantles and the grain cores.}              
\label{tab0bis}
\centering 
\renewcommand{\footnoterule}{}  
\begin{tabular}{ccc}     
\hline\hline       
 Species & \multicolumn{2}{c}{Abundances [$\chi$]} \\
          & Mantles & Cores \\ 
\hline                    
   H$_2$O & 7.25$\times$10$^{-5}$$^{\mathrm{(a)}}$ & $\ldots$ \\
   CH$_3$OH & $\sim$10$^{-6}$$^{\mathrm{(b)}}$ & $\ldots$ \\
   Si/SiO & $\sim$10$^{-8}$$^{\mathrm{(b)}}$ & $\sim$3.6$\times$10$^{-5}$$^{\mathrm{(c)}}$ \\
\hline                  
\end{tabular}
\begin{list}{}{}
\item[$^{\mathrm{(a)}}$] From \citet{whi91}.
\item[$^{\mathrm{(b)}}$] From \citet{jim05}.
\item[$^{\mathrm{(c)}}$] From \citet{and89} and \citet{snow96}.
\end{list}
\end{table}
%

The comparison of the model predictions with observations (Secs.$\,$5 
and 6) requires the consideration of the radial velocity of the 
preshock gas (ambient cloud gas) relative to the observer, $v_{cl}$. 
This velocity, the radial velocity of the emission 
measured by the observer, $v_{LSR}$, and the velocity of the neutral fluid 
as measured in the frame of the ambient medium, $v_n$, are related by:

   \begin{eqnarray}
      v_{LSR} = v_{cl}+v_n \label{vlsr}
   \end{eqnarray}

In Appendix A, we also give the equations for 
$v_n$, $v_i$ and $n_n$ within the frame of the shock (see Eqs.$\,$\ref{velsf} 
and \ref{denssf}) that will be used in Sec.$\,$4.1 to validate this parametric 
approximation.

\section{Sputtering of Grains}

In this section, we describe the sputtering of grains
produced by collisions with H$_2$ and He and other heavy
atomic and molecular species such as C, O, Si, Fe and CO (see Appendix
B for the full explanation of the method). 
Although we consider that most
silicon is locked into the olivine grain cores, we assume 
that a small fraction of this element is also present within the icy water
mantles ($q_m$=1.4$\times$10$^{-4}$;
  see below). CH$_3$OH has been also considered as another  
constituent of the icy mantles. 

%
\begin{table*}
\caption{Input parameters for the C-shock profiles shown in
    Figs.$\,$\ref{shst3} and \ref{shst2}.}             
\label{tab2}
\centering 
\renewcommand{\footnoterule}{}  
\begin{tabular}{ccccccccccc}     
\hline\hline       
\textit{v}$_s$ & \textit{n(H$_2$)} & \textit{B$_0$}
 & $\chi$$_e$ & \textit{z$_n$} & \textit{z$_i$} & 
\textit{z$_T$} & $z_0$ & \textit{a$_T$} 
& $\Delta$ & \textit{T}$_{n,max}$ \\ 
(km$\,$s$^{-1}$) & (cm$^{-3}$) & ($mG$) & & (cm) & (cm) & (cm) & (cm) &
(K$^{1/6}$cm$^{-1}$) & (cm) & (K) \\
\hline                    
   40 & 10$^4$ & 0.14 & 7$\times$10$^{-8}$ & 7.0$\times$10$^{15}$ & 
   1.0$\times$10$^{15}$ & 1.8$\times$10$^{15}$ & 8.0$\times$10$^{15}$ &
   9.0$\times$10$^{-16}$ & 5$\times$10$^{16}$ & 2000 \\
   35 & 10$^8$ & 14 & 7$\times$10$^{-10}$ & 6.0$\times$10$^{13}$ & 
   2.0$\times$10$^{13}$ & 1.3$\times$10$^{13}$ & 0.0 & 1.3$\times$10$^{-13}$
   & 2$\times$10$^{14}$ & 1700 \\
   
\hline                  
\end{tabular}
\end{table*}
%

   \begin{figure*}
   \includegraphics[angle=270, width=0.85\textwidth]{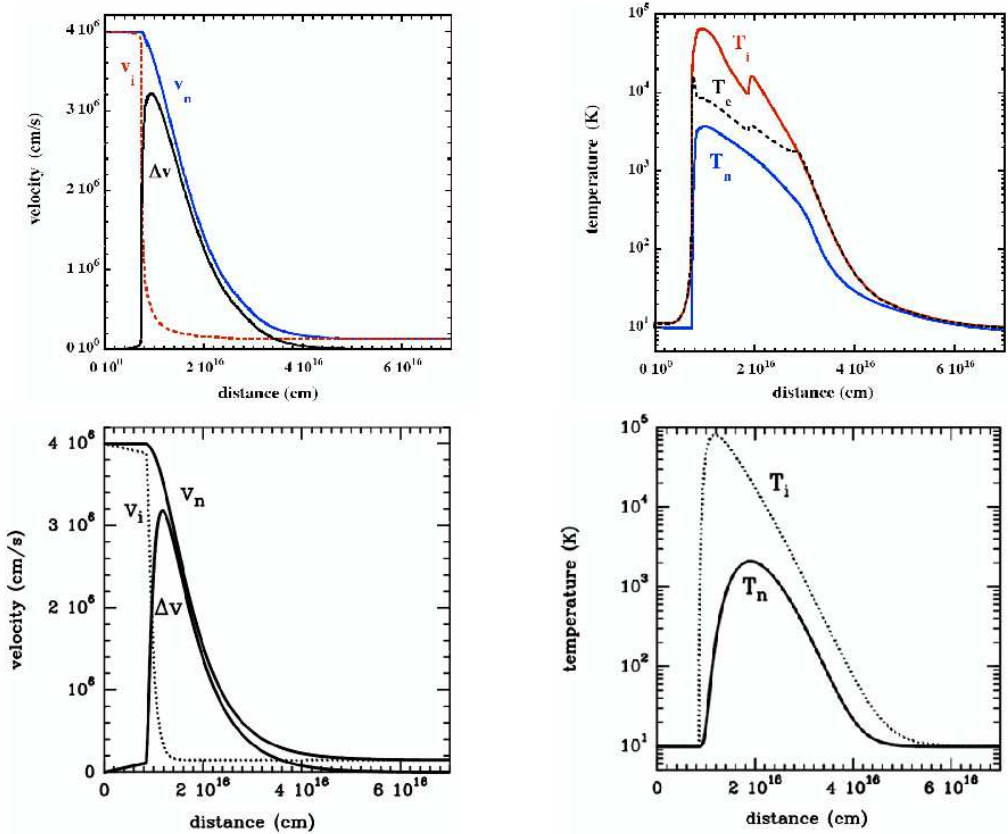}
   \caption{Comparison between our approximation of $v_n$, $v_i$,
     $\Delta v$, $T_n$ and $T_i$ (lower panels), and the MHD shock structure 
     calculated as in \citet[][upper panels]{flo03} for a shock with 
     $v_s$=40$\,$km$\,$s$^{-1}$, $n_0$=10$^4$$\,$cm$^{-3}$ and $B_0=100\mu G$
     (Flower, private communication). Shock parameters are
     given in Table$\,$\ref{tab2} and velocities are in the shock frame (see Appendix A for the definition of $v_n$ and $v_i$).} 
              \label{shst3}%
    \end{figure*}

\subsection{Sputtering of the Grain Mantles}

To study the sputtering of the grain mantles, we have followed the
procedure described by \citet{cas97}.  
The sputtering rate per unit volume and grain (Eq.$\,$\ref{dNdt} 
in Appendix B) has been derived by averaging the sputtering 
yield at low energies (Eq.$\,$\ref{Y0}) over a velocity-shifted 
Maxwellian distribution characterized by $T_n$ and $v_d$. 
The surface binding energy $U_0$ of the water mantles is of
0.53$\,$eV \citep[][]{tie94}. The projectile 
masses $m_p$ are 2, 4, 12, 16, 28, 56 
and 28$\,$amu for H$_2$, He, C, O, Si, Fe and CO, respectively. 
The target mass $M_t$ is considered to be 18$\,$amu which 
corresponds to the molecular mass 
of H$_2$O. The initial fractional abundances of He, C, O, Si, Fe
  and CO, relative to atomic hydrogen, are shown in Table$\,$\ref{tab0}. 
We assume that these abundances remain constant throughout the
dissipation region of the shock. 
The volume density of grains, $n_g$, 
is derived by considering a gas-to-dust mass ratio of $\sim$100, a constant
grain radius of 0.1$\,$$\mu$m and a density of the grain core material 
of 3.5$\,$g$\,$cm$^{-3}$ \citep[most of the volume of a grain is filled
by the silicate core;][]{cas97}. 
The total sputtering rate for H$_2$O, CH$_3$OH and silicon are
calculated with Eqs.$\,$(\ref{dNtoth2o}), (\ref{dNtotmet}) and (\ref{dNtotSi}). 
The total volume densities of these species are finally estimated 
with Eqs.$\,$(\ref{nagua}), (\ref{nmet}) and (\ref{nSi}).  
Since the amount of material contained within the grain mantles 
is limited, we assume that the maximum abundances of silicon 
and CH$_3$OH ejected from the mantles are those of SiO and 
CH$_3$OH measured in the low velocity gas of the L1448-mm outflow 
\citep[$\sim$10$^{-8}$ and $\sim$10$^{-6}$, 
respectively; see Table$\,$\ref{tab0bis} and][]{jim05}.
From the observations, and assuming a H$_2$O abundance of 
$\sim$7.25$\times$10$^{-5}$ \citep[Table$\,$\ref{tab0bis}
  and][]{whi91}, we can derive the fraction of silicon, $q_m$, and  
CH$_3$OH, $r_m$, present within the water mantles as
$q_m=\chi(SiO)/\chi(H_2O)$=1.4$\times$10$^{-4}$ and 
$r_m=\chi(CH_3OH)/\chi(H_2O)$=1.4$\times$10$^{-2}$. 
Although $q_m$ and $r_m$ constitute free parameters, we have fixed 
their values for comparison purposes with the L1448-mm outflow 
(Secs.$\,$5 and 6).  

\subsection{Sputtering of the Grain Cores}

For the sputtering of the cores, we have used different approaches to calculate
the sputtering produced by collisions with H$_2$, and by collisions
with He, C, O, Si, Fe and CO. We assume that olivine
(MgFeSiO$_4$) is the main form of silicates in the cores.
In the case of H$_2$, we calculate the angle-averaged sputtering yield as 
in Sec.$\,$3.1 (see Eq.$\,$B.2 in the Appendix B), but considering a 
surface binding energy $U_0$=5.70$\,$eV for the silicate cores \citep{tie94}. 
For the sputtering agents He, C, O, Si and Fe, 
we have used the sputtering yields for olivine calculated 
by \citet[][see Eq.$\,$\ref{Ymay}]{may00}. Since CO has the same 
projectile mass as Si, we assume that the sputtering yield of CO is
similar to that of Si \citep{fie97,may00}.  
The sputtering threshold energies $E_{th}$ used for the projectiles 
are 73$\,$eV for He, 48$\,$eV for C, and 47$\,$eV for O, Si, Fe and CO 
\citep[Table$\,$4 of][]{may00}. To take into account the
projection effects in the production of silicon within the
shock, we have included the factor 1/cos$^2$$\theta$ in the impact
energy $E_p$ of the colliding particle (Eq.$\,$\ref{Ymay}), where 
$\theta$ is the inclination angle of the outflow with 
respect to the line of sight. For the L1448-mm outflow (see
Secs.$\,$5 and 6), $\theta$ is $\sim$70$^{\circ}$ \citep{gir01}. 
As for the mantles, the total sputtering rate for silicon is determined
by Eq.$\,$(\ref{dNtotSi}), and (Eq.$\,$\ref{nSi}) calculates the volume density 
of silicon ejected from the grain cores. In the case of H$_2$, we
  also need to consider that the probability for a silicon
atom to be injected into the gas phase from an olivine molecule, as
opposed to a Mg, Fe or O atom, $q_c$, is 0.2 
\citep[][]{cas97}. However, for the rest of the colliding particles, we
assume that $q_c$=1 since this probability has been already taken into
account in the calculations of the sputtering yields of
\citet{may00}. Practically all
silicon ($\sim$99.97\%) is locked into the grain cores with an abundance of
$\sim$3.6$\times$10$^{-5}$ \citep[see
  Table$\,$\ref{tab0bis};][]{and89,snow96}.   

\section{Results}

\subsection{Validation of the Approximation of the Physical 
Structure of the C-shock}

To validate the parametric approximation of the C-shock physical 
structure of Sec.$\,$2, in Fig.$\,$\ref{shst3} we show the
comparison between the C-shock profile calculated as in the 
recent MHD models of \citet[][]{flo03}, and the profiles 
of $v_n$, $v_i$, $\Delta v$, $T_n$ and $T_i$ derived in the shock frame 
through our approach (see Appendix A for details). In Fig.$\,$\ref{shst2}, 
we directly compare our results with the MHD shock structure 
obtained by \citet{kau96}. The values of $z_n$, $z_i$,
$z_T$, $z_0$ and $a_T$ chosen to reproduce these shock profiles 
(see below for the estimation of the input parameters), and the values
of the magnetic field, $B_0$, fractional ionization, $\chi$$_e$, 
shock length scale, $\Delta$, and maximum 
temperature of the neutral fluid, $T_{n,max}$, are shown in 
Table$\,$\ref{tab2}. 

From Fig.$\,$\ref{shst3}, it is clear that,
although some differences do exist between the approximation 
and the MHD shock modelling at moderate preshock densities 
(Flower, private communication),
the general behaviour of $v_n$, $v_i$,
$\Delta v$, $T_n$ and $T_i$, qualitatively mimics the physical 
structure of C-shocks. In particular, the velocity decoupling 
between the ion and neutral fluids in the magnetic precursor, 
and the initial {\it delay} in the switch on of the neutral 
heating at this stage, are well reproduced by our approximation. 
Note that the agreement between the profiles of $v_n$ and
$v_i$ is excellent, giving a reliable prediction of $\Delta v$ which
is a key parameter in the sputtering yield calculation (see
Appendix B). In order to fit the delay 
of the heating of the neutrals at the magnetic precursor stage, 
we need to impose $b_T$$\geq$6. 

   \begin{figure}
   \centering
   \includegraphics[angle=270, width=0.5\textwidth]{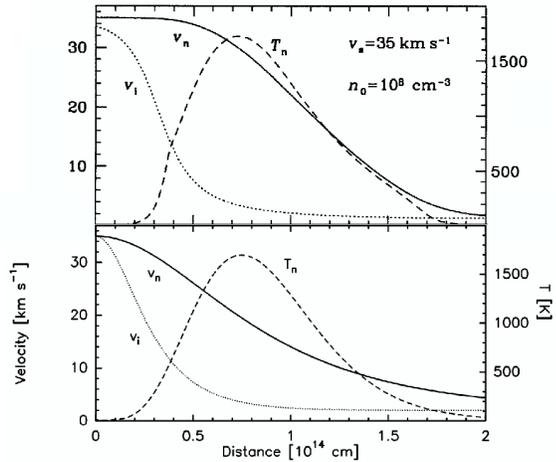}
   \caption{Comparison between the profiles of $v_n$, $v_i$ and $T_n$
   derived with our approximation (lower panel) and the MHD shock structure 
   obtained by \citet{kau96} for a shock with
   $v_s$=35$\,$km$\,$s$^{-1}$ and $n_0$=10$^8$$\,$km$\,$s$^{-1}$
   (upper panel). Parameters are given in Table$\,$\ref{tab2} and 
   velocities are in the shock frame.}
              \label{shst2}%
    \end{figure}

   \begin{figure}
   \centering
   \includegraphics[angle=0, width=0.45\textwidth]{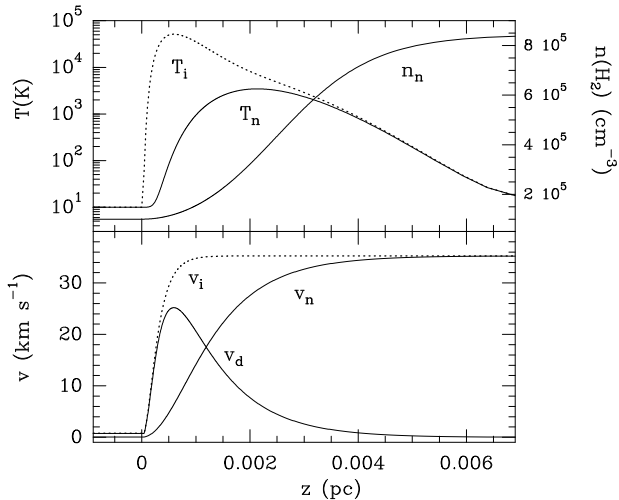}
   \caption{C-shock physical structure obtained with 
     Eqs.$\,$(1), (3) and (4) for 
     $v_s$$\,$=$\,$40$\,$km$\,$s$^{-1}$,
     $n_0$$\,$=$\,$10$^5$$\,$cm$^{-3}$ and
     $T_0$$\,$=$\,$10$\,$K. Velocities are in the frame co-moving with
     the preshock gas. The magnetic precursor length is of
     $\Delta z$$\sim$0.0005-0.001$\,$pc=1.5-3.0$\times$10$^{15}$$\,$cm.}
              \label{shst1}%
    \end{figure}

For higher preshock densities, Fig.$\,$\ref{shst2} shows that 
the ion-neutral velocity decoupling in the magnetic precursor,
is not as accurately reproduced as in Fig.$\,$\ref{shst3} for the
moderate density case \citep{flo03}. We should mention, however, that 
the MHD treatment of C-shocks for high density
regions, is still rather simplistic and therefore, uncertain
\citep[see][]{pil90,pil94,fal03,cha06}.  
Since the existing perpendicular shock models may not start 
having problems until the preshock density exceeds 
10$^6$$\,$cm$^{-3}$ \citep[see e.g.][]{pil90}, 
in the following we will restrict our study to the moderate density
case (from 10$^4$ to 10$^6$$\,$cm$^{-3}$).  

In Table~\ref{tab1}, we summarize the different values of $z_n$,
$z_i$, $z_T$ and $a_T$, chosen to reproduce the physical structure
of a sample of C-shocks with velocities of 
10$\leq$$v_s$$\leq$40$\,$km$\,$s$^{-1}$ and initial 
H$_2$ densities of 10$^4$$\leq$$n_0$$\leq$10$^6$$\,$cm$^{-3}$. 
$z_n$ and $z_i$ have been estimated by considering that 
$v_n$ is 0.999$\,$$v_s$ at $\Delta$,  
and by assuming a $z_n$$/$$z_i$ ratio of
$\sim$9$/$2. We note that slightly higher (factor of 1.5) $z_n/z_i$ ratios 
are required to reproduce the results of \citet{flo03} with a new treatment
of the coupling between the neutral and the charged fluids (Flower, private
communication). However, for consistency,
we will hereafter use the $z_n/z_i$ ratio of $\sim$9/2, since it well reproduces the
results of \citet{flo96} and \citet{drd83}, for which \citet{dop03} accordingly 
give an estimate of the shock length scale, $\Delta$. In any case, the results obtained
with both values of $z_n/z_i$ do not differ by more than 15\%.

The parameters $a_T$ and $z_T$ have been derived by
assuming that $b_T$=6 and $z_0$=0$\,$cm. 
The magnetic field, $B_0$, and fractional ionization of
the gas, $\chi_e$, have been calculated as in \citet{drd83}, 
and the shock length scale, $\Delta$, as in \citet{dop03}. 
The estimated Alfv\'en velocity is of $v_A$=2.18$\,$km$\,$s$^{-1}$, and 
$v_0$ typically ranges from 3.1 to 4.7$\,$km$\,$s$^{-1}$ for the cases 
considered in Table$\,$4 (see Appendix A for the calculation of $v_A$ and $v_0$).
We note that the shock length scales derived for every initial 
H$_2$ density of Table$\,$\ref{tab1}, are of the same order of magnitude
as the ion-neutral coupling lengths determined by 
\citet[][see Fig.$\,$1 in this work]{kau96}. 
The maximum temperature of the neutral fluid, $T_{n,max}$, has been
estimated from the results of \citet{drd83}.    

In Fig.$\,$\ref{shst1}, we show a representative profile of a C-shock
with $v_s$=40$\,$km$\,$s$^{-1}$, $n_0$=10$^5$$\,$cm$^{-3}$ and $T_0$=10$\,$K.
As expected in the
  frame co-moving with the preshock gas, the ion and neutral fluids
  are at rest at the beginning of the shock, and their final
  velocities in the far downstream gas are of $\sim
  v_s-v_0$ \citep[see][]{drd83}. 
The initial delay of the heating of
the neutrals gives the magnetic precursor length, which is of 
$\Delta z$$\sim$0.0005-0.001$\,$pc$\sim$1.5-3.0$\times$10$^{15}$$\,$cm
(Fig.$\,$\ref{shst1}). While the maximum value of 
the temperature of the ions is correlated with the 
maximum value of $v_d$ (see Fig.$\,$\ref{shst1}), 
the neutrals show their maximum 
temperature at $v_n$$\sim$0.85$\,$$v_s$ 
\citep[derived by assuming H$_2$O cooling and $\alpha$$_c$=1.5 in
  Eq.$\,$18 of][]{smi90}, which is consistent with the results of 
\citet[][see Fig.$\,$\ref{shst2}]{kau96}.  
 
From the recent results of time-dependent shock modelling, one may
consider that our assumption of steadiness for the C-shock could not be
valid to describe the time evolution of the sputtering of grains. 
These models indeed show that a J-type component is a natural
feature in the far downstream gas of the C-shock (near the {\it piston}) 
for time-scales of
$\leq$10$^3$-10$^4$$\,$yr, for which the steady state 
is finally attained \citep{chi98,les04}. However, as shown
in Sec.$\,$4.2, the evolutionary stages relevant to the main injection 
of the material contained in the icy mantles and in the grain
cores are those of the magnetic precursor which, independently 
on the age of the shock, can be described by the steady
state profile of C-shocks \citep[][]{chi98,les04}. 

%
\begin{table*}
\caption{Input parameters for a sample of C-shocks.}             
\label{tab1}
\centering 
\renewcommand{\footnoterule}{}  
\begin{tabular}{ccccccccccc}     
\hline\hline       
\textit{v}$_s$ & $v_0$$^{\mathrm{a}}$ & \textit{n(H$_2$)} & \textit{B$_0$}$^{\mathrm{b}}$
 & $\chi$$_e$$^{\mathrm{b}}$ & \textit{z$_n$} & \textit{z$_i$} & 
\textit{z$_T$}$^{\mathrm{c}}$ & \textit{a$_T$}$^{\mathrm{c}}$ 
& $\Delta$$^{\mathrm{d}}$ & \textit{T}$_{n,max}$$^{\mathrm{e}}$ \\ 
(km$\,$s$^{-1}$) & (km$\,$s$^{-1}$) & (cm$^{-3}$) & ($\mu$G) & & (cm) & (cm) & (cm) &
(K$^{1/6}$cm$^{-1}$) & (pc) & (K) \\
\hline                    
   20 & 3.8 & 10$^4$ & 140 & 7$\times$10$^{-8}$ & 1.4$\times$10$^{16}$ & 
   3.2$\times$10$^{15}$ & 5.0$\times$10$^{15}$ & 2.9$\times$10$^{-16}$
   & 0.024 & 900 \\
   40 & 4.7 & 10$^4$ & 140 & 7$\times$10$^{-8}$ & 2.8$\times$10$^{16}$ & 
   6.2$\times$10$^{15}$ & 1.1$\times$10$^{16}$ & 1.5$\times$10$^{-16}$
   & 0.048 & 2200 \\ \hline
   10 & 3.1 & 10$^5$ & 450 & 2$\times$10$^{-8}$ & 7.7$\times$10$^{14}$ & 
   1.7$\times$10$^{14}$ & 2.0$\times$10$^{14}$ & 5.8$\times$10$^{-15}$ 
   & 0.0012 & 300 \\
   20 & 3.8 & 10$^5$ & 450 & 2$\times$10$^{-8}$ & 1.4$\times$10$^{15}$ & 
   3.2$\times$10$^{14}$ & 5.0$\times$10$^{14}$ & 2.8$\times$10$^{-15}$ 
   & 0.0024 & 800 \\ 
   30 & 4.3 & 10$^5$ & 450 & 2$\times$10$^{-8}$ & 2.1$\times$10$^{15}$ & 
   4.7$\times$10$^{14}$ & 8.0$\times$10$^{14}$ & 2.0$\times$10$^{-15}$
   & 0.0036 & 2000 \\
   40 & 4.7 & 10$^5$ & 450 & 2$\times$10$^{-8}$ & 2.8$\times$10$^{15}$ & 
   6.2$\times$10$^{14}$ & 1.1$\times$10$^{15}$ & 1.6$\times$10$^{-15}$
   & 0.0048 & 4000 \\ \hline
   20 & 3.8 & 10$^6$ & 1400 & 7$\times$10$^{-9}$ & 1.4$\times$10$^{14}$ & 
   3.2$\times$10$^{13}$ & 5.0$\times$10$^{13}$ & 2.8$\times$10$^{-14}$
   & 2.4$\times$10$^{-4}$ & 800 \\
   40 & 4.7 & 10$^6$ & 1400 & 7$\times$10$^{-9}$ & 2.8$\times$10$^{14}$ & 
   6.2$\times$10$^{13}$ & 1.1$\times$10$^{14}$ & 1.6$\times$10$^{-14}$
   & 4.8$\times$10$^{-4}$ & 4000 \\
\hline                  
\end{tabular}
\begin{list}{}{}
\item[$^{\mathrm{a}}$] Calculated with Eq.$\,$\ref{v0} (see Appendix A) and 
assuming $v_A$=2.18$\,$km$\,$s$^{-1}$. 
\item[$^{\mathrm{b}}$] Estimated using Eqs.$\,$(62) and (63) of \citet{drd83}.
\item[$^{\mathrm{c}}$] Calculated considering that $b_T$=6 and $z_0$=0$\,$cm. 
\item[$^{\mathrm{d}}$] Derived as in \citet{dop03} and assuming
  that $n_{0,i}$$/$$n_H$$\,$$\sim$10$^{-6}$.
\item[$^{\mathrm{e}}$] Taken from Figs.$\,$8b and 9b of \citet{drd83} for
$n_0$=10$^4$$\,$cm$^{-3}$, and $n_0$=10$^5$ and 10$^6$$\,$cm$^{-3}$ 
respectively.
\end{list}
\end{table*}
%

\subsection{Sputtered Silicon Abundances: Injection and
     Saturation Times.}  

\begin{figure*}
   \centering
   \includegraphics[angle=270, width=1.0\textwidth]{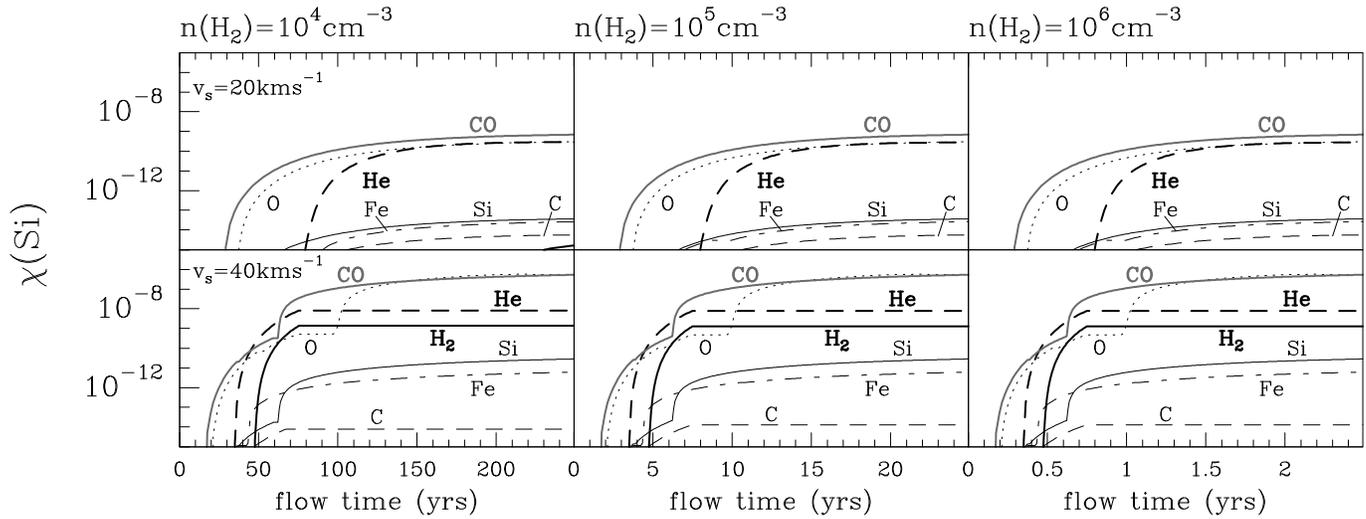}
   \caption{Evolution of the abundance of elemental silicon 
     ejected from grains by the impact with H$_2$, He, C, O,
     Si, Fe and CO, for initial H$_2$ densities of 10$^4$, 10$^5$ and
     10$^6$$\,$cm$^{-3}$ and shock velocities of 20 and
     40$\,$km$\,$s$^{-1}$.} 
              \label{plot2}%
    \end{figure*}

We now include the evolutionary profiles of $v_n$, $v_i$, $T_n$, $T_i$ and
$n_n$ from Sec.$\,$4.1 in the sputtering equations of the
Appendix B to calculate the silicon abundances
ejected from the mantles and from the cores. 
Fig.$\,$\ref{plot2} shows the silicon abundances 
ejected from grains as a function of the flow time for
several H$_2$ gas densities and shock velocities.
The abundances of sputtered silicon do not practically change 
with the initial density of the gas, which is clearly in agreement with the 
results of \citet{cas97}. However, as expected from the strongly
dependence of the sputtering rate on the maximum value of $v_d$ \citep[see
  Eq.$\,$\ref{dNdt} and][]{pin97}, the silicon abundance is drastically 
enhanced by increasing shock velocities. 
From Fig.$\,$\ref{plot2}, we also note that the time-scales are 
progressively reduced by nearly a factor of 10
as we increase the H$_2$ density from 10$^4$ to 10$^5$ and
10$^6$$\,$cm$^{-3}$. This is consistent with the fact that the flow
time, $t$, is inversely proportional to the density \citep[the
cooling time-scales roughly vary as $n_i$$^{-1}$, where $n_i$ is 
proportional to the density; see][]{chi98,les04}.   

\begin{figure*}
   \centering
   \includegraphics[angle=270, width=1\textwidth]{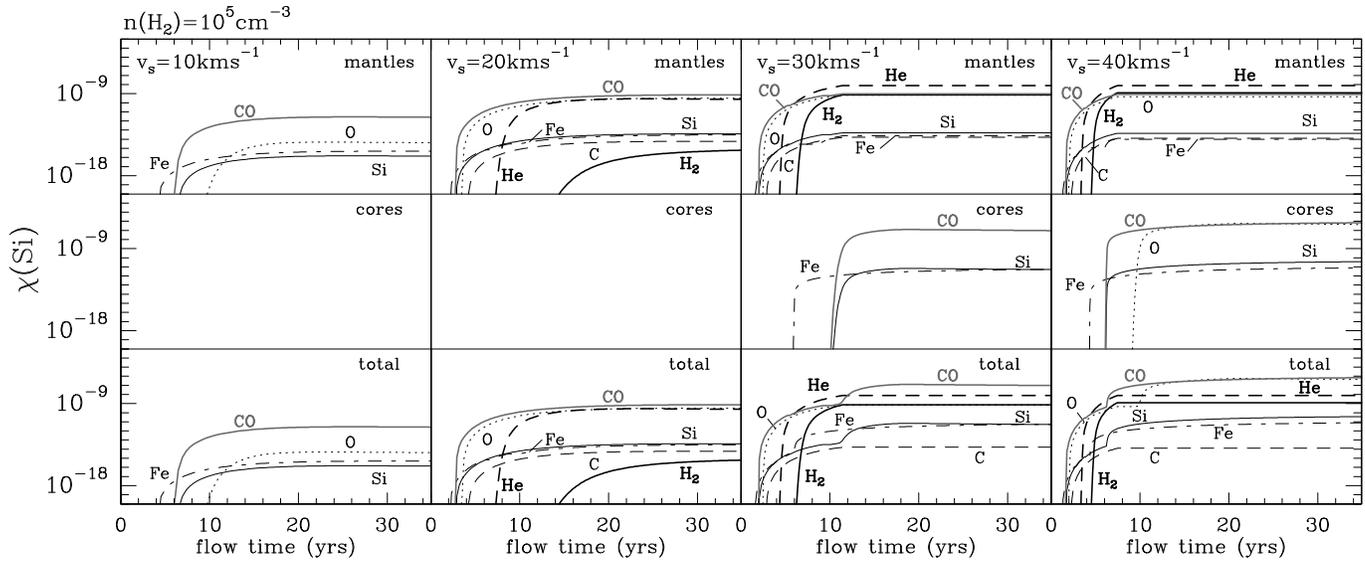}
   \caption{Abundance of elemental silicon ejected from grains by the
     impact with H$_2$, He, C, O, Si, Fe and CO, for shock velocities of
     10, 20, 30 and 40$\,$km$\,$s$^{-1}$ and a H$_2$ gas density 
     of 10$^5$$\,$cm$^{-3}$. For each shock velocity,
     we show the individual production of silicon from the
     mantles (upper panels), from the cores (middle panels), 
     and the total production of silicon from grains (lower panels).} 
              \label{plot3}%
    \end{figure*}

In Fig.$\,$\ref{plot3}, we show the products of the sputtering 
of the mantles and of the cores for an initial density of 
10$^5$$\,$cm$^{-3}$ and for shock velocities of 10, 20, 30
and 40$\,$km$\,$s$^{-1}$.
The sputtering of the grain mantles by collisions with 
H$_2$ and He (in bold lines in Fig.$\,$\ref{plot3}) exactly corresponds to that
previously calculated by \citet{cas97}. Although the fractional
abundance of the heavy species is orders of
magnitude smaller than that of H$_2$ and He (see Sec.$\,$3.1), it is
clear that the heavy atoms and CO sputter the mantles much more efficiently 
than H$_2$ or He for low shock velocities (Fig.$\,$\ref{plot3}). 

The high efficiency of these heavy species as sputtering agents is 
also shown by the \textit{injection}, $t_{inj}$, and \textit{saturation} 
times, $t_{sat}$, of Table$\,$\ref{tab3}.
We define $t_{inj}$ as the time for which the gas phase silicon abundance, 
relative to H$_2$, exceeds 10$^{-20}$ (i.e. the lower limit of
Fig.$\,$\ref{plot3}); and $t_{sat}$ as the time for which the difference 
(in the logarithmic scale) of the silicon abundance between two
consecutive time steps $t_{i+1}$ and $t_i$ (see Appendix A) is
$|log_{10}[\chi(m)]_{i+1}-log_{10}[\chi(m)]_i|$$<$0.1. Table$\,$\ref{tab3} also shows 
the injection and saturation times due to the contribution of 
all colliding particles. We note that these 
times are a factor of $\sim$100 smaller than the
typical dynamical ages of young molecular outflows like L1448-mm
($\sim$1000$\,$yr), but are roughly of the same order of magnitude as the
time-scales derived for the young shocks found in this outflow
\citep[$\leq$90$\,$yr;][]{gir01}. 
From Table$\,$\ref{tab3}, we find that the injection and saturation times of
H$_2$ and He are larger than those of the heavy
atoms and of CO. 

%
\begin{table*}
\caption{Injection and saturation times for the grain mantles and the grain cores 
for a medium with an initial H$_2$ density of $n_0=10^5\,cm^{-3}$.} 
\label{tab3}
\centering 
\renewcommand{\footnoterule}{}  
\begin{tabular}{ccccccccc|cccccccc}     
\hline\hline       
 & \multicolumn{8}{c}{{\bf Mantles}} & \multicolumn{8}{c}{{\bf Cores}}
 \\ \hline
 & \multicolumn{8}{c}{\textit{v}$_s$ (km$\,$s$^{-1}$)} &
 \multicolumn{8}{c}{\textit{v}$_s$ (km$\,$s$^{-1}$)} \\
& \multicolumn{2}{c}{10} & \multicolumn{2}{c}{20} &
 \multicolumn{2}{c}{30} & \multicolumn{2}{c}{40} 
& \multicolumn{2}{c}{10} & \multicolumn{2}{c}{20} &
 \multicolumn{2}{c}{30} & \multicolumn{2}{c}{40}  \\
\hline
& Inj. & Sat. & Inj. & Sat. & Inj. & Sat. & Inj. & Sat.
& Inj. & Sat. & Inj. & Sat. & Inj. & Sat. & Inj. & Sat. \\                  
   H$_2$ & $\ldots$ & $\ldots$ & 14.6 & 16.3 & 6.3 & 8.2 & 4.6  & 6.0 &
$\ldots$ & $\ldots$ & $\ldots$ & $\ldots$ & $\ldots$ & $\ldots$ &
     $\ldots$ & $\ldots$ \\
   He    & $\ldots$ & $\ldots$ & 7.4 & 10.1 & 4.4 & 6.3  & 3.4  & 5.0
   & $\ldots$ & $\ldots$ & $\ldots$ & $\ldots$ & $\ldots$ & $\ldots$ &
     $\ldots$ & $\ldots$ \\
   C     & $\ldots$ & $\ldots$ & 4.4  & 6.4 & 2.9 & 4.9  & 2.3  & 4.1
   & $\ldots$ & $\ldots$ & $\ldots$ & $\ldots$ & $\ldots$ & $\ldots$ &
     $\ldots$ & $\ldots$ \\
   O     & 9.8 & 14.4 & 3.7  & 6.0 & 2.5 & 4.7  & 2.1  & 3.9
   & $\ldots$ & $\ldots$ & $\ldots$ & $\ldots$ & $\ldots$ & $\ldots$ &
     9.2 & 10.4 \\
   Si    & 6.9  & 10.5  & 2.9  & 5.7  & 2.1 & 4.3  & 1.8  & 3.6
   & $\ldots$ & $\ldots$ & $\ldots$ & $\ldots$ & 10.5 & 11.9 &
     6.2  & 6.6 \\
   Fe    & 4.5     &  8.6    & 2.3  & 5.1  & 1.8 & 3.5  & 1.3  & 2.6
   & $\ldots$ & $\ldots$ & $\ldots$ & $\ldots$ & 5.9 &  6.5 &
     4.3 & 4.8 \\
   CO    & 6.5     &  10.5    & 2.9  & 5.7  & 2.1 & 4.3  & 1.8  & 3.6
   & $\ldots$ & $\ldots$ & $\ldots$ & $\ldots$ & 10.2 & 11.9 & 
     6.1 & 6.6  \\
  {\bf all} & {\bf 4.5} & {\bf 10.5} & {\bf 2.3}  & {\bf 5.7} & {\bf
 1.8} & {\bf 4.4}  & {\bf 1.3}  & {\bf 4.6} & {\bf $\ldots$} & {\bf
 $\ldots$} & {\bf $\ldots$} & {\bf $\ldots$} & {\bf 5.9} & {\bf 11.9} &
     {\bf 4.3} & {\bf 6.6} \\
\hline                  
\end{tabular}
\begin{list}{}{}
\item[{\tiny NOTE} .--] The injection and saturation times are given in yr.
\end{list}
\end{table*}
%

For the heavy colliding particles, CO seems to dominate the 
sputtering of the icy water mantles in low velocity shocks.  
Although Fe initiates the grain sputtering (Fe has the smallest injection 
and saturation times; see Table$\,$\ref{tab3}), its low fractional 
abundance prevents large enhancements of silicon by the impact 
with this element ($\leq$3$\times$10$^{-14}$ for $v_s\leq$20$\,$km$\,$s$^{-1}$; Fig.$\,$\ref{plot3}). 
On the contrary, collisions with CO (whose injection and saturation times are 
very similar to those of Si but whose initial abundance is 4 
orders of magnitude larger than that of Si) produce the main
injection of silicon from the mantles. The saturation times for 
CO are indeed very similar to those derived for all
colliding particles at low shock velocities (see Table$\,$\ref{tab3}). 

The abundance of silicon sputtered by collisions with CO for
$v_s$$\leq$10$\,$km$\,$s$^{-1}$ is very low
($\sim$10$^{-12}$; Fig.$\,$\ref{plot3}). However, at 
slightly higher shock velocities ($v_s=$20$\,$km$\,$s$^{-1}$), 
this abundant molecule can eject from the mantles considerable 
amounts of this element ($\sim$10$^{-9}$). Averaging this silicon
abundance over the dissipation region (shock length scale of
$\sim$7$\times$10$^{15}$$\,$cm; see Tab.$\,$\ref{tab1}), 
we estimate that the total column density of silicon injected into the gas
phase in a 20$\,$km$\,$s$^{-1}$-shock 
is of $\sim$10$^{12}$$\,$cm$^{-2}$. While shock
velocities of $v_s$$\geq$25$\,$km$\,$s$^{-1}$ were required 
to obtain Si/SiO column densities of $\geq$10$^{12}$$\,$cm$^{-2}$ in 
\citet{may00}, we find that the inclusion of silicon as a 
minor constituent of the grain mantles reduces the
sputtering {\it threshold velocity} by, at least, 
$|\sim$5$\,$km$\,$s$^{-1}|$ in our model with respect to previous
results. This velocity threshold is even reduced by
$|\sim$10$\,$km$\,$s$^{-1}|$ compared to the results of \citet{cas97}.  


It is also interesting to note that CO can also generate
silicon abundances of $\sim$2$\times$10$^{-11}$
at the very early stages of low velocity shocks ($t\leq$10$\,$yr; 
see cases with $v_s$=20 and 30$\,$km$\,$s$^{-1}$ in
Fig.$\,$\ref{plot3}). As discussed in Sec.$\,$5, these results
could explain the detection of SiO abundances of $\sim$10$^{-11}$
associated with the narrow SiO emission observed in the young
shocks of the L1448-mm outflow. 

For shocks with $v_s\geq$30$\,$km$\,$s$^{-1}$, 
He plays an important
role in the sputtering of the mantles. Note that the saturation times
for all colliding particles at these shock velocities, slightly
deviate from those of CO due to the increasing efficiency of He to erode 
the grain mantles (Table$\,$\ref{tab3} and Fig.$\,$\ref{plot3}). 
Almost all silicon within the mantles 
($\sim$8$\times$10$^{-9}$) is injected into the gas phase by
collisions with He for shock velocities of $v_s\sim$30$\,$km$\,$s$^{-1}$. 

For the sputtering of the grain cores, 
only collisions with O, Si, Fe and CO are efficient enough
to destroy the cores. The injection of silicon into the gas
phase from the cores occurs for shock velocities
$v_s$$\geq$30$\,$km$\,$s$^{-1}$, which is consistent with the
  results of \citet{may00}. For
$v_s$$\,$=$\,$40$\,$km$\,$s$^{-1}$, only $\sim$3\% of the total amount
of silicon locked into the grain cores, is released into the gas
phase (abundance of $\sim$10$^{-6}$; see
Fig.$\,$\ref{plot3}). As for
the mantles, and although Si has the same injection and saturation times as
those of CO (both are assumed to have similar sputtering
properties for the cores; see Sec.$\,$3.2), CO is the main sputtering 
agent of the grain cores since its initial fractional abundance clearly
exceeds that of Si. 

\section{Comparison with Observations: The SiO Abundances}  

If we now assume that silicon is rapidly 
oxidized into SiO \citep{pin97} or that SiO is directly released from 
grains \citep{mar92}, we can directly compare our predictions of
the silicon abundance ejected from grains by the sputtering, 
with the SiO abundances observed in very young bipolar outflows like
in L1448-mm. 

\begin{figure}
   \centering
   \includegraphics[angle=0, width=0.45\textwidth]{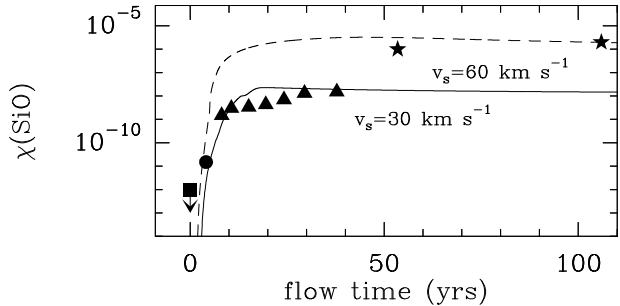}
   \caption{Predicted SiO abundances 
   within a 30$\,$km$\,$s$^{-1}$- and a
   60$\,$km$\,$s$^{-1}$-shock as a function of the flow time (see shock
   parameters in Table$\,$\ref{tab4}). Observational SiO abundances derived
   in the ambient gas (filled square), the precursor component (filled
   circle), the moderate velocity gas (filled triangles) and the high
   velocity regime (filled stars) found in the L1448-mm outflow 
   \citep[][]{mar92,jim05}, are also shown. The black arrow
   indicates an upper limit to the SiO abundance. The observed flow times 
   have been derived from Eq.$\,$\ref{time} of Sec.$\,$2 (see text and Appendix A 
   for details).}
              \label{plot4}%
    \end{figure}

In Fig.$\,$\ref{plot4}, we show the SiO abundances measured
for the different velocity components detected in this outflow 
\citep[ambient gas, the shock precursor component, the moderate 
velocity gas and the high velocity gas;][]{jim05,mar92} as a function 
of the flow time. For the ambient and precursor
components, we have assumed central radial velocities of $v_{LSR}=$4.7 and
5.2$\,$km$\,$s$^{-1}$, respectively \citep[][]{jim04}.
Subtracting the ambient cloud velocity of $v_{cl}=$4.7$\,$km$\,$s$^{-1}$ 
characteristic of the molecular emission in L1448-mm  
from the central radial velocities of these components 
(see Eq.$\,$\ref{vlsr} in Sec.$\,$2), we obtain
flow velocities of $v_n=$0 and 0.5$\,$km$\,$s$^{-1}$, which correspond to
flow times of $t$=0 and 4.0$\,$yrs for the ambient gas and the precursor
component in a 30$\,$km$\,$s$^{-1}$-shock (Eq.$\,$\ref{time} of Sec.$\,$2;
see also Eq.$\,$\ref{trapz} in Appendix A for the details on the computation of 
the flow times). For the moderate velocity gas, 
we have considered velocity intervals of 1$\,$km$\,$s$^{-1}$-width between 
6 and 18$\,$km$\,$s$^{-1}$
\citep[][]{jim05}. The SiO abundances
for the ambient, precursor and moderate velocity gas
have been derived assuming optically thin emission and excitation
temperatures of $\sim$10-15$\,$K \citep[][]{req07}. 
For the high velocity gas, the SiO abundances of
$\sim$10$^{-6}$ and 2$\times$10$^{-6}$ (Fig.$\,$\ref{plot4})
have been taken from 
Tab.$\,$6 in \citet{mar92} for the velocity ranges of
$-$50$\,$km$\,$s$^{-1}$$\leq$$V_{LSR}$$\leq$$-$40$\,$km$\,$s$^{-1}$
and $-$60$\,$km$\,$s$^{-1}$$\leq$$V_{LSR}$$\leq$$-$50$\,$km$\,$s$^{-1}$. 
The flow time associated with each velocity 
interval is again estimated from their central radial velocities, $v_{LSR}$, after
subtracting the ambient cloud velocity of the L1448-mm outflow of 
$v_{cl}$=4.7$\,$km$\,$s$^{-1}$ (Eq.$\,$\ref{vlsr}). 

%
\begin{table*}
\caption{Parameters of the C-shock models that best fit the SiO and CH$_3$OH 
observational data shown in Figs.$\,$\ref{plot4} and \ref{plot5}.}             
\label{tab4}
\centering 
\renewcommand{\footnoterule}{}  
\begin{tabular}{cccccccccc}     
\hline\hline       
$v_s$ & $n$(H$_2$) & $B_0$ & $\Delta$ & $T_{n,max}$ & $z_n$ & $z_i$ & 
$z_T$ & $a_T$ & $z_n/z_i$ \\ 
(km$\,$s$^{-1}$) & (cm$^{-3}$) & ($\mu G$) & (pc) & (K) & (cm) & (cm) & (cm) &
(K$^{1/6}$cm$^{-1}$) & \\
\hline     

30 & 10$^5$ & 450 & 0.0036 & 2000 & 2.1$\times$10$^{15}$ &
5.0$\times$10$^{14}$ & 8.0$\times$10$^{14}$ & 2.0$\times$10$^{-15}$ &
4.2 \\

60 & 10$^5$ & 450 & 0.0072 & 6000 & 4.1$\times$10$^{15}$ &
1.0$\times$10$^{15}$ & 1.1$\times$10$^{15}$ & 1.8$\times$10$^{-15}$ &
4.0 \\ \hline

\end{tabular}
\end{table*}

Fig.$\,$\ref{plot4} also
shows the silicon abundances, as a function of time, predicted 
by our model for the shocks which  
best fit the observational SiO data (with $v_s$=30$\,$km$\,$s$^{-1}$ and
$v_s$=60$\,$km$\,$s$^{-1}$). 
The shock parameters used to reproduce the physical structure of these 
shocks are shown in Table$\,$\ref{tab4}. 

From Fig.$\,$\ref{plot4}, we note that the sputtering produced by the 
propagation of a 30$\,$km$\,$s$^{-1}$-shock perfectly matches the 
evolutionary trend of SiO to be enhanced from the 
ambient to the moderate velocity gas observed in L1448-mm \citep{jim05}. 
The progressive erosion of the icy mantles by the
sputtering with CO, reproduces the SiO abundances observed 
in the ambient gas ($\leq$10$^{-12}$; filled square), in the precursor 
component ($\sim$10$^{-11}$; filled circle), and in the 
moderate velocity gas (from $\sim$10$^{-9}$ to $\sim$10$^{-8}$; 
filled triangles in Fig.$\,$\ref{plot4}). This suggests that the puzzling
narrow SiO line detected toward the young shocks of L1448-mm
can be explained by the recent erosion of the grain mantles 
containing a small fraction of Si/SiO, at the early 
stages of low velocity shocks.     

To fit the SiO abundances
measured in the high velocity gas, a velocity shock with 
$v_s$=60$\,$km$\,$s$^{-1}$ is needed to sputter $\sim$9\% of the
silicon contained within the olivine cores and increase
the predicted silicon abundance up to a few 10$^{-6}$ (filled
stars in Fig.$\,$\ref{plot4}). This shock velocity is clearly
in excess of the critical velocities of C-shocks
\citep[$v_{crit}\sim$40-50$\,$km$\,$s$^{-1}$; see][]{drd83,smi90}. 
\citet{bou02} and \citet{cab04} have recently shown that
the actual shock velocity limit can be increased to 
$v_{crit}$$\sim$100$\,$km$\,$s$^{-1}$ for moderate densities and high
magnetic fields. However, we cannot rule out the possibility that a J-type
component would be the responsible for the large SiO abundances
observed in the high velocity gas of L1448-mm.   

Given the fact that the L1448-mm outflow shows variability in its
  high velocity SiO emission, an alternative scenario would involve the 
  presence of two different shocks at two different evolutionary stages  
that would coexist within the single-dish beam of the SiO observations 
\citep{jim05}.

\section{Sputtered CH$_3$OH and H$_2$O Abundances.}  

\begin{figure}
   \centering
   \includegraphics[angle=0, width=0.45\textwidth]{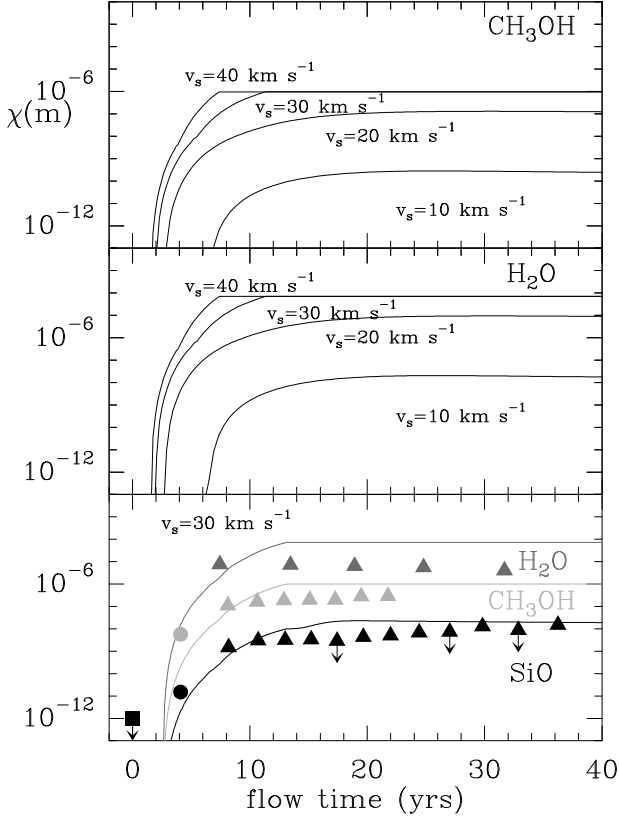}
   \caption{{\it Upper and middle panels:} Predictions of the
   sputtered abundances of CH$_3$OH and H$_2$O as a function of the
   flow time, for a sample of 
   C-shocks with $v_s$=10, 20, 30 and 40$\,$km$\,$s$^{-1}$ and
   $n_0$=10$^5$$\,$cm$^{-3}$. 
   {\it Lower panel:} SiO, CH$_3$OH and H$_2$O abundances ejected from
   the grain mantles for the 30$\,$km$\,$s$^{-1}$-shock of Table$\,$\ref{tab4} 
   as a function of time. We also show the
   SiO, CH$_3$OH and H$_2$O abundances measured in the ambient gas
   (square), the precursor component (circles) and the moderate
   velocity gas (triangles) of the L1448-mm outflow 
   \citep{jim05,ben02}. Vertical arrows
   indicate upper limits to the SiO abundance.} 
              \label{plot5}%
    \end{figure}

In addition to SiO, CH$_3$OH and H$_2$O are also expected 
to be largely enhanced in outflow regions \citep[see e.g.][]{drd83,kau96}.
In Fig.$\,$\ref{plot5} (upper and middle panels), we
show the predicted abundances of CH$_3$OH and H$_2$O as a function of
the flow time, for a sample of shocks with $v_s$=10, 20, 30 and 
40$\,$km$\,$s$^{-1}$ (see Table$\,$\ref{tab1}). 
The impact with CO injects CH$_3$OH and H$_2$O abundances 
as large as $\sim$10$^{-7}$ and $\sim$10$^{-5}$, respectively, 
in low velocity shocks (see cases with $v_s$=20$\,$km$\,$s$^{-1}$; 
Fig.$\,$\ref{plot5}). These abundances are even enhanced to up to
$\sim$10$^{-6}$ for CH$_3$OH, and to $\sim$10$^{-4}$ for H$_2$O, in shocks
with only $v_s$=30$\,$km$\,$s$^{-1}$. 

We can now compare the predicted abundances of CH$_3$OH and H$_2$O 
with those observed in the young
shocks of the L1448-mm outflow. Fig.$\,$\ref{plot5} (lower panel) 
shows the SiO and CH$_3$OH abundances observed in the ambient gas 
(square), the precursor component (circles) and the
moderate velocity gas (triangles) of this outflow 
\citep[][]{jim05}, as a function of the flow time. In this figure, we
also show the abundances of ortho-H$_2$O derived from the line profile of the
1$_{10}$$\rightarrow$1$_{01}$ transition 
measured by SWAS \citep{ben02}.
Since the velocity resolution of these observations \citep[
$\sim$1$\,$km$\,$s$^{-1}$;][]{ben02} is 
lower than those of the SiO and CH$_3$OH spectra 
\citep[$\sim$0.14$\,$km$\,$s$^{-1}$; see][]{jim05}, we have only estimated
the abundances of ortho-H$_2$O for the moderate velocity regime 
(dark grey triangles in Fig.$\,$\ref{plot5}). As in Sec.$\,$5, the flow times
associated with these abundances have been inferred subtracting the 
ambient cloud velocity in L1448-mm of $v_{cl}$=4.7$\,$km$\,$s$^{-1}$ from 
the central velocities, $v_{LSR}$, 
of the observed line profiles of these molecules 
(see Eq.$\,$\ref{vlsr}).
The SiO, CH$_3$OH and H$_2$O abundances have been derived assuming 
optically thin emission as in \citet[][]{jim05} for SiO and CH$_3$OH, 
and as in \citet[][]{neu00} for H$_2$O (Eq.$\,$1 in this work). 
The abundance of SiO (black line), CH$_3$OH (light grey line) and
H$_2$O (dark grey line) in gas phase 
generated by the sputtering of the grain mantles in the 
30$\,$km$\,$s$^{-1}$-shock of
Sec.$\,$5 (see Table$\,$\ref{tab4}) are also shown in Fig.$\,$\ref{plot5}.  

As for SiO, the abundance of sputtered CH$_3$OH is in agreement with
that derived in the precursor component within one order of magnitude, and
with those measured in the moderate velocity gas within a factor of $\leq$5
(see Fig.$\,$\ref{plot5}). We would like to stress that we have only
argued the total amount of material in the grain mantles. 
This implies that the progressive enhancement
of the SiO and CH$_3$OH abundances observed for velocities of 
$\leq$20$\,$km$\,$s$^{-1}$ toward L1448-mm \citep{jim05}, is naturally 
explained by the presence of a single shock with
$v_s$=30$\,$km$\,$s$^{-1}$. 
In the case of ortho-H$_2$O, however, the abundances of this molecule 
in the moderate velocity gas differ by more than a factor of 
10 from those predicted by our model. This could be due either to the
assumption of optically thin emission for the ortho-H$_2$O
1$_{10}$$\rightarrow$1$_{01}$ transition \citep[which could
have underestimated the ortho-H$_2$O abundances;][]{ben02}, or to a
beam dilution effect. Note that the SWAS beam is of $\sim$240$''$, i.e., 8
times larger than the 30$\,$m beam of the SiO $J$=2$\rightarrow$1
observations \citep{jim05}. This leads to a {\it corrected} 
ortho-H$_2$O abundance of some $\sim$10$^{-4}$, which is consistent
with our model predictions of Fig.$\,$\ref{plot5}.

\section{Conclusions}

In this work, we have presented a parametric model that mimics 
the steady state profile of the physical parameters of C-shocks. 
The simplicity of this model has allowed, for the
first time, the detailed analysis of the time evolution of the sputtering of
the grain mantles and the grain cores in regions with recent 
outflow activity. Although this approximation does not include 
detailed MHD modelling, we have shown that it 
can be used as an efficient tool to interpret the time dependent 
evolution of the abundances of typical
shock tracers like SiO, CH$_3$OH, H$_2$O or NH$_3$ in young molecular 
outflows, where transient phenomena are expected to play an important
role. The assumption of steadiness of the shock can be
applied for these young objects,
since only the early stages of the shock evolution 
(characterized exclusively by the interaction of the magnetic
precursor) are relevant for the process of the sputtering of dust
grains.   

To calculate the sputtering of the grain mantles and the grain cores, 
we have assumed that silicon and
methanol are minor constituents of the water mantles, and that olivine
is the main form of silicates within the grain cores.  
In spite of the low fractional abundance of heavy atoms (C, O, Si and
Fe) and molecules (CO) relative to H$_2$ or He in molecular dark clouds, 
these species have been also considered as sputtering agents. The
relatively high abundance of CO with respect to the rest of 
heavy colliding particles, and its large sputtering yield, makes the
sputtering by CO very efficient. Collisions with this molecule
can eject a considerably large fraction of silicon in
the mantles for shocks with only
$v_s\sim$20$\,$km$\,$s$^{-1}$. This implies a reduction by
$|\sim$5-10$\,$km$\,$s$^{-1}|$ of the
{\it threshold velocity} of the sputtering with respect to other models. 

By comparing the evolution of the abundances of SiO, CH$_3$OH and H$_2$O
predicted by our model, with the abundances derived 
toward the different velocity regimes found in the young shocks of
the L1448-mm outflow, we find that two different shocks (with 
$v_s$=30$\,$km$\,$s$^{-1}$ and $v_s$=60$\,$km$\,$s$^{-1}$) are 
needed to reproduce the measured abundances. 
The progressive enhancement of SiO and CH$_3$OH
observed from the ambient gas up to moderate velocities,
is consistent with the mantle erosion produced by a 
single 30$\,$km$\,$s$^{-1}$-shock. We find that the 
SiO abundance of $\sim$10$^{-11}$ associated with the very narrow SiO
emission detected in this outflow, can be explained
as an early product of the sputtering of the mantles by CO in low
velocity shocks if the Si/SiO abundance in the grain mantles is
$\sim$10$^{-4}$ with respect to water. 
The disagreement between the predicted and the
derived ortho-H$_2$O abundances is probably due either to the
assumption of optically thin emission for the ortho-H$_2$O
1$_{10}$$\rightarrow$1$_{01}$ line observed by SWAS, or to a
beam dilution effect. 

The approximation presented in this work, not only will furnish an
input for more comprehensive MHD models of the shock structure in
young molecular outflows, but will allow to perform direct
comparisons with the molecular line profiles 
observed toward these regions. These comparisons will be 
presented in a future paper.


\begin{acknowledgements}

We are indebted to
Prof. D. R. Flower for his comments on the C-shock and sputtering
theory and for kindly providing the MHD shock structure shown
in Fig.$\,$\ref{shst3}. We acknowledge Dr. M. Kaufman for letting 
us use his MHD shock profile shown in Fig.$\,$\ref{shst2}.
We would like to thank Dr. A. Asensio-Ramos 
for the help provided during the development of the code, and
Dr. F. Daniel for his useful comments on the manuscript. We also
acknowledge an anonymous referee for helping us to significantly improve the paper.
This work was supported by the Spanish MEC through projects
number AYA2003-02785-E, ESP2004-00665 and ESP2007-65812-C02-01, 
and by the ``Comunidad de
Madrid'' Government under PRICIT project S-0505$/$ESP-0277 (ASTROCAM).
PC acknowledges support from the Italian Ministry of Research and
University within a PRIN project.

\end{acknowledgements}

\appendix

\section{The C-shock Structure in the Shock Frame}

The steady-state velocity profiles of the ion and neutral
fluids, $v_i^*$ and $v_n^*$, within the frame of the shock, are approximated as:

   \begin{eqnarray}
      v_{n,i}^* & = & v_0 + \frac{\left(v_s-v_0\right)}{cosh\left[(z-z_0)/z_{n,i}\right]} \label{velsf}
   \end{eqnarray}
  
where $z$ is the spatial coordinate and $v_s$ is the velocity of the shock.
$z_0$, $z_i$ and $z_n$ are input parameters that govern the velocity decoupling between 
the ions and the neutrals (see Secs.$\,$2 and 4.1), and $v_0$ is required 
to avoid infinite compression of the far 
downstream gas (see Eq.$\,$\ref{denssf} and Sec.$\,$2). This speed 
only depends on the speed of the shock, $v_s$, and the Alfv\'en speed, 
$v_A$ (see Eq.$\,$\ref{v0} below). From 
Eq.$\,$\ref{velsf}, it is clear that the ion and neutral fluids initially move 
with velocities $v_s$ and are progressively decelerated to 
$v_0$ (see Figs.$\,$\ref{shst3} and \ref{shst2}).
   
Assuming that the thermal pressure of the fluid is negligible compared to 
the magnetic and the dynamic pressures, we can estimate the magnitude of $v_0$ 
in the downstream gas from:

   \begin{eqnarray}
      \rho_0 v_s^2 + \frac{B_0^2}{8\pi} =  \rho v_0^2 + \frac{B^2}{8 \pi} \label{lawec}
   \end{eqnarray}

where $\rho_0$ and $\rho$ are the preshock and postshock gas 
mass densities, and $B_0$ 
and $B$ are the magnitude of the magnetic field in the preshock and postshock 
regimes, respectively. Considering that the compression of the magnetic field 
is given by $B = B_0 (v_s/v_0)$ and that the Alfv\'en speed, $v_A$, 
is defined as $v_A^2 = B_0^2/(4 \pi \rho)$, Eq.$\,$\ref{lawec} becomes:

   \begin{eqnarray}
      v_0^2 = \frac{v_s^2}{\rho/\rho_0} + \frac{v_A^2}{2}\left[1-\left(\frac{\rho}{\rho_0}\right)^2\right] \label{v02}
   \end{eqnarray}
 
From the principle of mass conservation, the fluid density 
in the postshock gas, $\rho$, is derived as $\rho=\rho_0 (v_s/v_0)$. 
If we now substitute the ratio $\rho/\rho_0$ 
in Eq.$\,$\ref{v02}, we finally obtain:

   \begin{eqnarray}
      v_0^2 \left[v_0^2-v_0v_s-\frac{v_A^2}{2}\right] & = & -\frac{v_A^2}{2} v_s^2 \label{v0}
   \end{eqnarray}
    
from which $v_0$ can be calculated. 

The particle density of the neutral fluid, $n_n$, is given by:

  \begin{eqnarray}
      n_n = n_0 v_s/v_n^* \label{denssf}
   \end{eqnarray}

and the time associated with the neutral fluid or flow time, $t$, in the 
frame of the shock is calculated as:

   \begin{eqnarray}
      t = \int \frac{dz}{v_n^*} \label{timesf}
   \end{eqnarray}

In each plane-parallel slab of material $i+1$ within the shock,  
the flow time, $t_{i+1}$, is calculated by using the 
trapezoidal method:

   \begin{eqnarray}
      t_{i+1} = t_{i}+\left(z_{i+1}-z_{i}\right)\left(\frac{f(z_{i})+f(z_{i+1})}{2}\right)\label{trapz}
   \end{eqnarray}

where $z_i$ and $z_{i+1}$ are the spatial coordinates for the slabs 
of gas $i$ and $i+1$, and $f(z_i)$ is defined as (see Eq.$\,$\ref{timesf}):

   \begin{eqnarray}
   f(z_i) = \frac{1}{v_n^*(z_i)} \label{fsf}
   \end{eqnarray}

The temperature of the ion and neutral fluids, $T_i$ and $T_n$, are 
estimated as in Sec.$\,$2.

\section{Sputtering of Grains. Silicon, CH$_3$OH and H$_2$O 
Fractional Abundances.}

The sputtering of grains has been calculated by considering different 
sputtering yields for the mantles and for the cores. 
The sputtering rate per unit
volume for a spherical target of radius $a$ moving with drift 
velocity $v_d$ through a Maxwellian neutral gas of temperature $T_n$ is
\citep[Eq.$\,$27 in][]{dra79}:


\begin{equation}
\label{cc}
\begin{array}{l}
\left[\frac{dn(m)}{dt}\right]_{grain} = \pi a^2n_p\left(\frac{8kT_n}{\pi m_p}\right)^{1/2} {_\times} \\
\\ 
\ \ \ \ \ \ \ \ \ \ \ \ \ \ \ \ \ \ \ \ \ \ _\times\frac{1}{s}\int_{x_{th}}^{\infty}dx\,x^2\,\frac{1}{2}\left[e^{-(x-s)^2}-e^{-(x+s)^2}\right]<Y(E)>_\theta
 \label{dNdt}
\end{array}
\end{equation}

where $n_p$ and $m_p$ are the number density and mass of the projectile,
respectively; and $s$ and $x$ are related to $v_d$, $T_n$ and the projectile
impact energy $E_p$ through  $s^2$$\,$=$\frac{m_pv_d^2}{2kT_n}$ 
and $E_p$$\,$=$\,$$x^2kT_n$. 
 
The angle-averaged sputtering yield at low energies $<Y(E)>_\theta$
for the mantles can be 
approximated by $<Y(E)>_\theta$$\approx$2$Y(E,\theta = 0)$ 
\citep[][]{dra95}, where the normal-incidence 
yield $Y(E,\theta = 0)$ is calculated as \citep[Eq.$\,$31; ][]{dra79}:

   \begin{equation}
      Y(E,\theta=0) = A \frac{(\varepsilon -
      \varepsilon_0)^2}{1+\left(\varepsilon/30\right)^{4/3}}, \, \, \,
      \varepsilon > \varepsilon_0 
   \label{Y0}
   \end{equation}
 
$A$ is a constant ($A\approx8.3\times10^{-4}$), and $\varepsilon$ 
and $\varepsilon_0$ are calculated as
$\varepsilon = \eta E_p/U_0$ and $\varepsilon_0=max[1,4\eta]$. $U_0$ is 
the binding energy (per atom or molecule) and $\eta$ 
is derived by doing $\eta = 4 \xi m_p M_t (m_p+M_t)^{-2}$, where 
$M_t$ is the target mass and $\xi$ is an 
efficiency factor which varies from 0.8 (for ices) to 1
\citep[for atomic solids;][]{dra79}. $x_{th}$, which is related to the 
threshold impact energy $E_{th}$, is finally calculated as
$x_{th}=\left(\frac{\varepsilon_0U_0}{\eta k T_n} \right)^{1/2}$.   

For the grain cores, we have used the sputtering yield calculated by 
\citet{may00} for the impact of atomic species on olivine cores. The 
sputtering yield is derived from: 

   \begin{equation}
      <Y(E)>_\theta = k_{s}\, exp\left[-\beta/(E_p-E_{th})\right]
   \label{Ymay}
   \end{equation}

where $k_{s}$, $\beta$ and $E_{th}$ (the sputtering threshold energy) 
are taken from Table$\,$4 in \citet{may00}. In this case, $x_{th}$ is
derived by doing $x_{th}=\left(\frac{E_{th}}{kT_n}\right)$.  

In each collision between projectile and grain, only a small fraction
of silicon, $q_m$, and CH$_3$OH, $r_m$, will be ejected from the mantles 
[$q_m$=1.4$\times$10$^{-4}$ and
$r_m$=1.4$\times$10$^{-2}$; see Sec.$\,$3.1]. Analogously, 
only a fraction of silicon, $q_c$, will be released from the cores
[$q_c$=0.2 for $H_2$ and $q_c$=1 for the rest of colliding particles; 
see Sec.$\,$3.2].
If we assume a grain density $n_g$, the total sputtering 
rate for H$_2$O, CH$_3$OH and silicon is:

   \begin{eqnarray}
      \left[\frac{dn(H_2O)}{dt}\right]_{tot}^{m} =n_g\,\left[\frac{dn(H_2O)}{dt}\right]_{grain}^{m} \label{dNtoth2o}
   \end{eqnarray}

   \begin{eqnarray}
      \left[\frac{dn(CH_3OH)}{dt}\right]_{tot}^{m} =n_g\,r_m\,\left[\frac{dn(CH_3OH)}{dt}\right]_{grain}^{m} \label{dNtotmet}
   \end{eqnarray}

  \begin{eqnarray}
      \left[\frac{dn(Si)}{dt}\right]_{tot}^{m,c}=n_g\,q_{m,c}\,\left[\frac{dn(Si)}{dt}\right]_{grain}^{m,c}\label{dNtotSi}
   \end{eqnarray}

where $m$ and $c$ denotes the sputtering rate for the 
mantles and the cores.

By using the Euler's algorithm, we calculate the total volume  
density of H$_2$O, CH$_3$OH and Si ejected from grains in each plane-parallel 
slab of material $i$ within the shock as: 

   \begin{eqnarray}
      n(H_2O)_{i+1} = n(H_2O)_{i}+\Delta t \left[\frac{dn(H_2O)}{dt}\right]_{tot,i}^m \label{nagua}
   \end{eqnarray}

   \begin{eqnarray}
      n(CH_3OH)_{i+1} = n(CH_3OH)_{i}+\Delta t \left[\frac{dn(CH_3OH)}{dt}\right]_{tot,i}^m \label{nmet}
   \end{eqnarray}

   \begin{eqnarray}
      n(Si)_{i+1} = n(Si)_{i}+\Delta t\left[\frac{dn(Si)}{dt}\right]_{tot,i}^{m,c}\label{nSi}
   \end{eqnarray}

where $\Delta t = (t_{i+1}-t_i)$. In this case, the flow time at the slab 
of material $i+1$, $t_{i+1}$, is numerically calculated as in Appendix A 
(see Eq.$\,$\ref{trapz}), but using the function $f(z_i)$ defined as (see Sec.$\,$2):

   \begin{eqnarray}
   f(z_i) = \frac{1}{v_s-v_n(z_i)} \label{f}
   \end{eqnarray}

The fractional abundance of silicon, H$_2$O and  CH$_3$OH, is finally derived 
by doing $\chiup(m)=n(m)/n(H_2)$ in each slab of material.

\end{document}